\documentclass[aps,twocolumn,pra,superscriptaddress,showpacs,10pt]{revtex4-1} 
\usepackage{setspace}
\usepackage{graphicx}
\usepackage{subfigure}
\usepackage{epsfig}
\usepackage{amsmath}
\usepackage{epsfig}

\begin{document}

\title{Driven dipole oscillations and the lowest energy excitations\\ of 
strongly interacting lattice bosons in a harmonic trap}

\author{K. He}
\affiliation{Department of Physics, The Pennsylvania State University,
University Park, Pennsylvania 16802, USA}
\affiliation{Department of Physics, Georgetown University, Washington, DC 20057, USA}
\author{J. Brown}
\affiliation{Department of Physics, Georgetown University, Washington, DC 20057, USA}
\author{S. Haas}
\affiliation{Department of Physics \& Astronomy, University of Southern 
California, California 90089, USA}
\affiliation{School of Engineering
and Science, Jacobs University Bremen, Bremen 28759, Germany}
\author{M. Rigol}
\affiliation{Department of Physics, The Pennsylvania State University,
University Park, Pennsylvania 16802, USA}

\begin{abstract}
We show that the analysis of the time evolution of the occupation of site and momentum modes 
of harmonically trapped lattice hard-core bosons, under driven dipole oscillations, allows 
one to determine the energy of the lowest one-particle excitations of the system 
in equilibrium. The analytic solution of a single particle in the absence of a lattice is used 
to identify which function of those time-dependent observables is best fit for the analysis, 
as well as to relate the dynamic response of the system to its single-particle spectrum. 
In the presence of the lattice and of multiple particles, a much richer and informative
dynamical response is observed under the drive.
\end{abstract}


\maketitle

\section{Introduction}\label{sect:intro}

Ultracold atomic gases in one-dimensional (1D) geometries exhibit a rich 
phenomenology \cite{cazalilla_citro_review_11} and display remarkable nonequilibrium 
phenomena \cite{kinoshita_wenger_06,trotzky_chen_12,gring_kuhnert_12}. They have 
been the center of much recent experimental and theoretical interest because of 
the possibility of controlling the potentials used to trap and manipulate these 
gases and studying their coherent dynamics \cite{bloch_dalibard_review_08}. 
For example, using optical lattices, experimentalists have accessed the  
strongly interacting Tonks-Girardeau regime in 1D bosonic systems 
\cite{kinoshita_wenger_04,paredes_widera_04} and examined their dynamics 
\cite{kinoshita_wenger_06}.

In addition to being of interest in their own right 
\cite{rigol_dunjko_07,polkovnikov_sengupta_review_11}, the dynamics of strongly 
correlated one-dimensional systems can be used to probe equilibrium properties 
not otherwise accessible. For example, the energy absorption rates obtained during 
the modulation of the amplitude \cite{iucci_cazalilla_06,kollath06_iucci_06a,orso_iucci_09} 
and phase \cite{tokuno_giamarchi_11} of an optical lattice have been used to gain insights into 
the spectrum of energy excitations in multiple phases of one-dimensional bosonic systems. 
Unfortunately, long simulation times and the need for independent calculations 
for each probed frequency have been a major obstacle for unbiased numerical studies 
of the lowest energy excitations in trapped lattice systems.

Here, we explore an alternative route that allows us to address those challenges. 
We examine the dynamics of site and momentum occupations of 1D lattice 
hard-core bosons (HCBs) under driven dipole oscillations. The time-dependent 
Hamiltonian of interest has the form
\begin{eqnarray}\label{hamil}
&&\hat{H}(t)=\hat{H}_0+\hat{H}_1(t),\nonumber\\
&&\hat{H}_0= -J\sum_{i=1}^{L-1}\left(\hat{b}_i^\dagger\hat{b}_{i+1}^{} + \text{H.c.}\right)
+V\sum_{i=1}^{L}\left(i-\frac{L+1}{2}\right)^2 \hat{n}_i,\nonumber \\
&&\hat{H}_1(t) = 2V A\sin(\omega't) \sum_{i=1}^{L} \left(i-\frac{L+1}{2}\right)\hat{n}_i,
\end{eqnarray}
where $\hat{b}_i^\dagger$ ($\hat{b}_i^{}$) is the creation (annihilation) operator 
of a HCB at site $i$ (satisfying the constraints $\hat{b}_i^{2}=\hat{b}_i^{\dagger2}=0$), 
$\hat{n}_i=\hat{b}_i^\dagger\hat{b}_i^{}$ is the site occupation operator,
$J$ is the nearest-neighbor hopping parameter, $V$ is the strength
of the harmonic trapping potential, $A$ is the amplitude of the driving, 
$\omega'$ is its frequency, and $L$ is the number of lattice sites ($A\ll L$). Note that 
the part of the Hamiltonian [Eq.~\eqref{hamil}] that gives the potential energy is the expansion, up to the 
linear term, of $V\sum_{i=1}^{L}\left[i-\frac{L+1}{2}+A\sin(\omega't)\right]^2 \hat{n}_i$.
As such, it can be generated in experiments by either directly adding a linear
time-dependent potential or by means of a small periodic displacement of the center of the 
trap; both generate dipole oscillations. In the absence of a drive, dipole oscillations
of bosons in optical lattices have already been studied experimentally \cite{fertig_ohara_05}
and theoretically \cite{polkovnikov_wang_04,rigol_rousseau_05,ruostekoski_isella_05,
rey_pupillo_05,pupillo_rey_06Dip,danshita_clark_09,montangero_fazio_09,danshita_13}.

We show that the parametric excitations due to the aforementioned driving and 
their signatures in the considered observables provide insight into the 
lowest-energy excitations of the {\it global} spectrum of the system. The exposition is 
organized as follows. In Sec.~\ref{sect:num}, we describe the numerical
approach used. In Sec.~\ref{sect:one}, we discuss the single particle solution,
followed by the general numerical analysis of the many-particle case in 
Sec.~\ref{sect:many}. The conclusions are presented in Sec.~\ref{sect:con}.

\section{Numerical approach}\label{sect:num}

The time evolution of the site and momentum occupations are computed
by mapping hard-core bosons onto noninteracting spinless fermions and 
using properties of Slater determinants as discussed in detail in 
Refs.~\cite{rigol_muramatsu_04,rigol_muramatsu_05a,rigol_muramatsu_05b}. This 
approach is exact, and the computation times involved scale polynomially with 
system size, which allows us to study large systems for long times. Since the
Hamiltonian \eqref{hamil} is time dependent (not the case in 
Refs.~\cite{rigol_muramatsu_04,rigol_muramatsu_05a,rigol_muramatsu_05b}), 
we use a second-order Trotter-Suzuki decomposition 
\cite{trotter_59,suzuki_76,deraedt_deraedt_83} to compute 
the time evolution of the wave function,
\begin{equation}
 |\Psi(t+\delta t)\rangle=e^{-\frac{i}{\hbar}\frac{\hat{H}_1(t+\frac{\delta t}{2})}{2}\delta t}
 e^{-\frac{i}{\hbar}\hat{H}_0 \delta t}e^{-\frac{i}{\hbar}
 \frac{\hat{H}_1(t+\frac{\delta t}{2})}{2}\delta t}|\Psi(t)\rangle,
\end{equation}
which introduces an error $O(\delta t^3)$ \cite{deraedt_deraedt_83}. 
$|\Psi(t+\delta t)\rangle$ can be efficiently calculated in our case because 
$e^{-\frac{i}{\hbar}\hat{H}_0 \delta t}$, being time independent, needs to be 
computed only once (it is done exactly by diagonalizing $\hat{H}_0$). This 
leaves the trivial computation of 
$e^{-\frac{i}{\hbar}\frac{\hat{H}_1(t+\frac{\delta t}{2})}{2}\delta t}$, from 
the already diagonal $\hat{H}_1(t)$, to be done at each time step. 

In our calculations, we consider $L=101$, $V/J=0.0036$, $A=1$, and 
$\delta t=0.005 \hbar/J$. At $t=0$, the system is taken to be in the ground state of 
$\hat{H}_0$, and we simulate the time evolution up to $t=5000 \hbar/J$. To assess 
the accuracy of the results, we computed the overlap between the wavefunctions obtained 
using the above value of $\delta t$ and twice that value, at the latest time simulated. 
For the maximal number of particles considered ($N_p=50$), the absolute value 
of that overlap is $0.99999992$. This gives us confidence in the high accuracy 
of our calculations.

\section{One-particle solution}\label{sect:one}

We start our study of the time evolution of the site occupancies and momentum 
distributions by analyzing their dynamics for a single particle under the proposed 
driving. In the presence of a lattice, the low-energy single-particle 
excitation spectrum of a system in which $V\ll J$, such as ours, resembles that of a 
harmonically trapped system in the continuum \cite{rigol_muramatsu_04b}. This means 
that under the assumption of a weak driving away from resonance (so that only the 
lowest-energy excitations in the lattice are involved), we can gain insights into 
this system by studying it in the continuum. The Schr\"odinger equation in this case reads
\begin{equation}\label{schro}
 \imath\hbar\frac{\partial\psi}{\partial t}=-\frac{\hbar^2}{2m}\frac{\partial^2\psi}{\partial x^2}
 +\frac{m\omega_0^2x^2}{2}\psi-m\omega_0^2Aa\sin(\omega't)x\,\psi,
\end{equation}
where $a$ is the lattice spacing [in Eq.~\eqref{hamil}, the amplitude of the driving was given 
in units of the lattice spacing], $m$ is the mass of the particle, and $\omega_0$ is the frequency
of the trapping potential. The latter two are related to the lattice parameters by the 
expressions $m=\hbar^2/2Ja^2$ and $\omega_0^2=4VJ/\hbar^2$. 

Equation~\eqref{schro} admits an exact analytical solution of the form (up to a constant prefactor) 
\cite{husimi_53}
\begin{equation}\label{hus}
 \psi(x,t)=\exp\left\{\frac{\imath}{2\hbar}\left[m\omega_0\alpha(t)x^2+
 2m\omega_0 x_0\beta(t)x+\hbar\gamma(t)\right]\right\},
\end{equation}
where $x_0=\sqrt{\hbar/m\omega_0}$ is the harmonic oscillator characteristic length and 
the dimensionless parameters $\alpha(t),\ \beta(t)$, and $\gamma(t)$ satisfy the equations
\begin{eqnarray}
\frac{d\alpha(t)}{dt}&=&-\omega_0 \alpha(t)^2-\omega_0, 
\quad \frac{d\gamma(t)}{dt}=\imath \omega_0 \alpha(t)-\omega_0\beta(t)^2,\nonumber\\
\frac{d\beta(t)}{dt}&=&-\omega_0\alpha(t)\beta(t)+\omega_0 B\sin(\omega't),
\end{eqnarray}
where $B=Aa/x_0$ is also a dimensionless parameter. Given our initial condition that 
$\psi(x,t=0)$ is the ground state of the harmonic oscillator, the above set of equations 
admits a straightforward solution; $\alpha(t)=\imath$. The expressions for $\beta(t)$ and 
$\gamma(t)$ are lengthy and not particularly informative (beyond telling us that the 
motion is periodic and that the condition for resonance is $\omega'=\omega_0$).
However, if we focus on the behavior of $\varrho(t)=\ln |\psi(x=0,t)|^2\propto
-\text{Im}[\gamma(t)]$, we obtain
\begin{equation}\label{log1}
 \varrho(t)=\varrho(0)-\frac{B^2 \omega_0^2}{(\omega_0^2 - \omega'^2)^2} 
 [\omega'\sin(\omega_0 t) - \omega_0 \sin(\omega' t)]^2,
\end{equation}
which is remarkably simple and has a frequency Fourier transform equal to a sum of Dirac 
$\delta$ functions at $\omega=0,\ \pm 2\omega_0,\ \pm 2\omega',\ \pm(\omega_0-\omega')$, 
and $\pm(\omega_0+\omega')$. Selecting $\omega'<\omega_0$, as we do in the following,
means that a Fourier transform will produce $\delta$ functions at the positive frequencies 
(i) $\omega=\omega_0\pm\omega'$, which allows us to identify the lowest excitation in 
the spectrum of the harmonic oscillator, (ii) $\omega=2\omega_0$, which allows us to 
identify the second lowest excitation, and (iii) $\omega=2\omega'$, which is related 
to the driving frequency.

Furthermore, the momentum Fourier transform of Eq.~\eqref{hus} has a simple form, 
whose expression at zero momentum (up to a constant prefactor) reads
\begin{equation}\label{husp}
 \psi(p=0,t)=\frac{1}{\sqrt{-\imath \alpha(t)}}\exp\left[-\frac{\imath}{2}
 \left(\frac{\beta(t)^2}{\alpha(t)}-\gamma(t)\right)\right].
\end{equation}
Since in our case $\alpha(t)=\imath$, we can define the quantity 
$\varsigma(t)=\ln |\psi(p=0,t)|^2\propto-\text{Im}[\imath\beta(t)^2+\gamma(t)]$, 
which reads
\begin{equation}\label{log2}
 \varsigma(t)=\varsigma(0)-\frac{B^2 \omega_0^2\omega'^2}{(\omega_0^2 - \omega'^2)^2} 
 [\cos(\omega_0 t)-\cos(\omega' t)]^2.
\end{equation}
The Fourier transform of $\varsigma(t)$ is a sum of Dirac $\delta$ functions 
at the same frequencies as those for $\varrho(t)$. It is important to notice that 
the functions $\varrho(t)$ and $\varsigma(t)$ are even functions 
of $t$. This time symmetry is expected because the reflection symmetry of 
the initial state about $x=0$ and $p=0$ means that the probability of finding 
the particle at $x=0$ or $p=0$ in the driven system must be independent of 
the sign in the last term in Eq.~\eqref{schro}. In order to simplify the 
exposition, in what follows we set $\hbar=1$.

\section{Many-particles in a lattice}\label{sect:many}

We now study numerically what happens in the presence of a lattice
as the number of particles $N_p$ in the trap is increased. Since this model can be 
mapped onto noninteracting spinless fermions \cite{lieb_shultz_61}, its spectrum 
of excitations coincides with that of the noninteracting fermions. The many-body 
ground state is created by occupying the lowest $N_p$ single-particle energy eigenstates, 
with energies $E(0)$ through $E(N_p-1)$ [$E(0)$ being the single-particle ground-state 
energy]. The first (one-particle) excitation corresponds to $E_1(N_p)=E(N_p)-E(N_p-1)$.
The second one corresponds to $E_2(N_p)=E(N_p+1)-E(N_p-1)$, which (particularly at low 
fillings) is nearly degenerate with $E'_2(N_p)=E(N_p)-E(N_p-2)$. Since a straightforward 
implementation of our approach does not resolve the difference between $E_2(N_p)$ and 
$E'_2(N_p)$, we treat them as one and only report $E_2(N_p)$. The next-lowest one-particle 
excitation is $E_3(N_p)=E(N_p+2)-E(N_p-1)$, which is nearly degenerate with 
$E'_3(N_p)=E(N_p+1)-E(N_p-2)$ and $E''_3(N_p)=E(N_p)-E(N_p-3)$ [again, we only report
$E_3(N_p)$], and so on. In the absence 
of a lattice all excitations would be multiples of $\omega_0$, but the lattice changes 
this dramatically. As discussed in Ref.~\cite{rigol_muramatsu_04b}, as $N_p$ increases, 
$E_1(N_p)$ decreases until it vanishes. At that point doubly degenerate eigenstates 
appear. They have zero weight over a growing region in the center of the trap and are 
related to the emergence of an $n_i=\langle \hat{n}_i\rangle=1$ insulator 
\cite{rigol_muramatsu_04b}. As a result, multiple properties of the lattice system 
are qualitatively different from those in the continuum 
\cite{rigol_muramatsu_04b,hooley_quintanilla_04,rey_pupillo_05,rigol_muramatsu_04c,rigol_muramatsu_05c}.

\begin{figure}[!t]
 \includegraphics[width=0.47\textwidth, angle=-0]{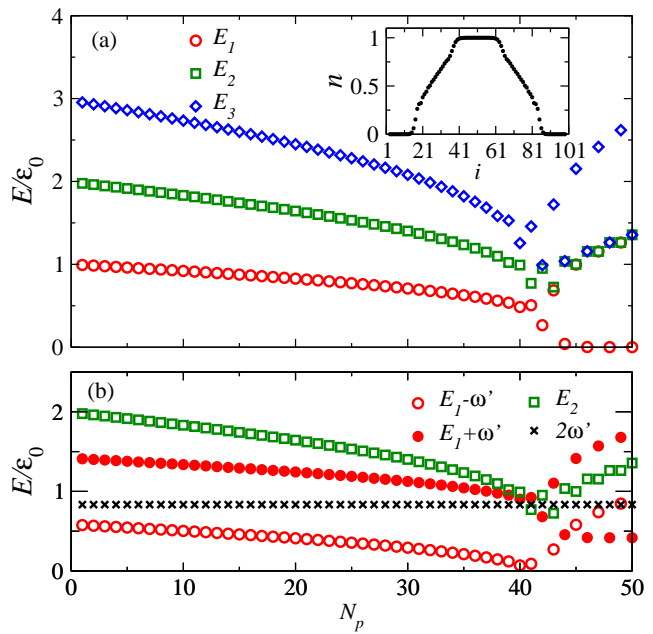}
 \vspace{-0.2cm}
\caption{(Color online) (a) Three lowest one-particle energy excitations as a function of 
the number of particles in the ground state. The inset shows the site 
occupations for $N_p=50$. (b) Expected frequencies for the largest 
response in the Fourier transform of $\ln[n_{i=51}(t)]$ and $\ln[m_{k=0}(t)]$, 
which follows from the prediction for one particle in the continuum while taking 
into account that the spectrum changes because of the presence of a lattice.}
 \label{fig:spect}
\end{figure}

In Fig.~\ref{fig:spect}(a), we show $E_1$, $E_2$, and $E_3$ vs $N_p$ for our system
at $t=0$. As expected, for small values of $N_p$, they are approximately equal to 
$\varepsilon_0 l$, where $\varepsilon_0=\omega_0$ and $l=1,\ 2,$ and 3, and 
decrease with increasing $N_p$. Degeneracies set in for $N_p\geq46$. At those fillings, 
a Mott insulator with $n_i=1$ can be seen in the ground-state site occupancies, 
as shown in the inset in Fig.~\ref{fig:spect}(a). Figure 1(b) depicts the frequencies 
at which the Fourier transform of the time evolution of the central site occupation 
and that of the zero momentum node should exhibit the largest response with increasing $N_p$, 
according to the single-particle results extended to account for the lattice effects
depicted in Fig.~\ref{fig:spect}(a). We take $\omega'=0.05J$ to be the driving 
frequency and $\omega_0=0.12J$ to be the trapping frequency.

In experiments with ultracold gases the momentum distribution function
$m_k$ can be determined in time-of-flight measurements, in which all 
confining potentials are turned off and the system is allowed to evolve 
freely \cite{cazalilla_citro_review_11,bloch_dalibard_review_08}, while 
the recent use of very high resolution optical imaging systems has made 
measuring site occupancies $n_i$ feasible 
\cite{bakr_gillen_09,bakr_peng_10,sherson_weitenberg_10,weitenberg_endres_11}. 
In what follows, we focus on the dynamics of those quantities under the drive.

\begin{figure}[!t]
 \includegraphics[width=0.47\textwidth, angle=-0]{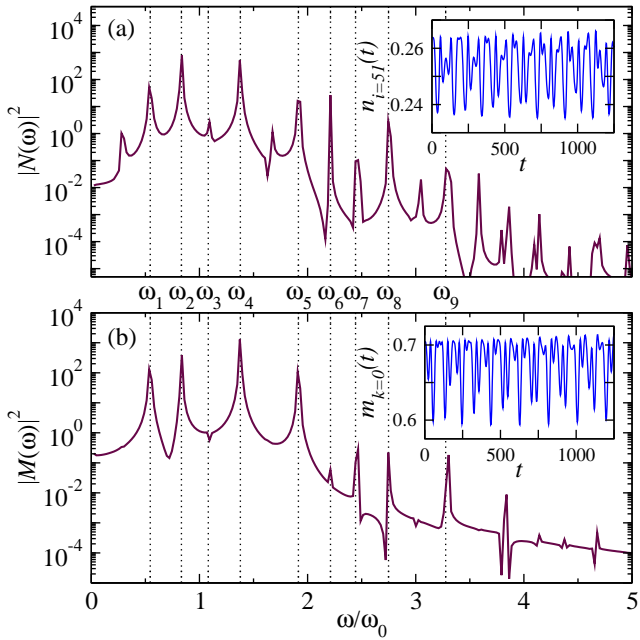}
  \vspace{-0.2cm}
\caption{(Color online) Fourier transform of the time evolution of the occupation of (a) the 
site at the center of the trap ($i=51$) and (b) the zero-momentum mode in a system with 
$N_p=5$ HCBs. The inset in each panel depicts the time evolution of the respective observable
($t$ is given in units of $J^{-1}$). Vertical dashed lines in the 
main panels indicate the most prominent frequencies highlighted in the Fourier transform of 
both observables. They correspond to $\omega_1=E_1-\omega'$, $\omega_2=2\omega'$, 
$\omega_3=E_2-2\omega'$, $\omega_4=E_1+\omega'$, $\omega_5=E_2$, 
$\omega_6=E_1+3\omega'$, $\omega_7=E_3-\omega'$, $\omega_8=E_2+2\omega'$, 
and $\omega_9=E_3+\omega$.}
 \label{fig:FT}
\end{figure}

In the insets in Fig.~\ref{fig:FT}, we show the time evolution of the occupation 
of the site at the trap center $n_{i=51}$ [inset in Fig.~\ref{fig:FT}(a)] 
and that of the zero momentum occupation [inset in Fig.~\ref{fig:FT}(b)]. They exhibit 
periodic dynamics in which multiple frequencies are involved, and at 
the shortest times $t>0$, both observables decrease as predicted 
by Eqs.~\eqref{log1} and \eqref{log2}. In our 
numerical calculations the observables are measured in intervals 
$\Delta t=2J^{-1}$. In addition, when computing the Fourier transforms of 
$\ln[n_{i=51}(t)]$ and of $\ln[m_{k=0}(t)]$, we only 
considered times in the interval $500 J^{-1}<t\leq 5000 J^{-1}$. By not taking 
into account results for earlier times, we reduce the effect of any transient 
behavior that may affect our results.

In the main panels in Fig.~\ref{fig:FT}, we show the Fourier transforms 
of $\ln[n_{i=51}(t)]$ [$N(\omega)$ in Fig.~\ref{fig:FT}(a)] and of $\ln[m_{k=0}(t)]$
[$M(\omega)$ in Fig.~\ref{fig:FT}(b)] for a system with $N_p=5$. For both observables, 
we find that the four most prominent peaks [better seen in $M(\omega)$ in 
Fig.~\ref{fig:FT}(b)] are at the frequencies $\omega_{1,4}=E_1\mp\omega'$, 
$\omega_2=2\omega'$, and $\omega_5=E_2$ as predicted by the analysis for one 
particle in the continuum. Note that the values of $E_1$ and $E_2$ are those
depicted in Fig.~\ref{fig:spect} for $N_p=5$ and were obtained from exactly
diagonalizing the Hamiltonian at $t=0$. They depart from the values $E_1=\omega_0$
and $E_2=2\omega_0$ expected in the continuum. 

In addition to those four frequencies, we find that others are also 
highlighted by the Fourier analysis as the number of particles 
is increased. The most prominent ones with signatures in both observables 
are $\omega_{3,8}=E_2\mp2\omega'$, $\omega_6=E_1+3\omega'$,
and $\omega_{7,9}=E_3\mp\omega'$. We note that $\omega_{7,9}$ are just
signatures of $E_3$ displaced by $\mp\omega'$, respectively, similar to 
$\omega_{1,4}$ for $E_1$. This means that $E_3$ can also be 
identified by analyzing the dynamics in the lattice. 
Our results also show that other replicas of $E_1$ can be 
found to be displaced by $(2l+1)$ multiples of $\omega'$ (where $l>0$ is an integer)
and those of $E_2$ can be found to be displaced by $2l$ multiples of $\omega'$. 
Hence, there is a pattern by which the 
frequencies of one-particle transitions that change the parity of the ground
state are displaced by odd multiples of $\omega'$, while the frequencies 
of the transitions that do not change the parity of the ground state are 
displaced by even multiples of $\omega'$ (including zero). Since we are 
computing the Fourier transforms of $\ln[n_{i=51}(t)]$ and of $\ln[m_{k=0}(t)]$, 
this pattern can be understood to be a consequence of the invariance 
of $n_{i=51}(t)$ and $m_{k=0}(t)$ under $t\rightarrow -t$, as discussed
for the one-particle case. Only specific combinations of periodic functions 
of $\omega t$ and $\omega' t$ appear to ensure that the 
resulting functions are even in time.

\begin{figure}[!t]
 \includegraphics[width=0.475\textwidth, angle=-0]{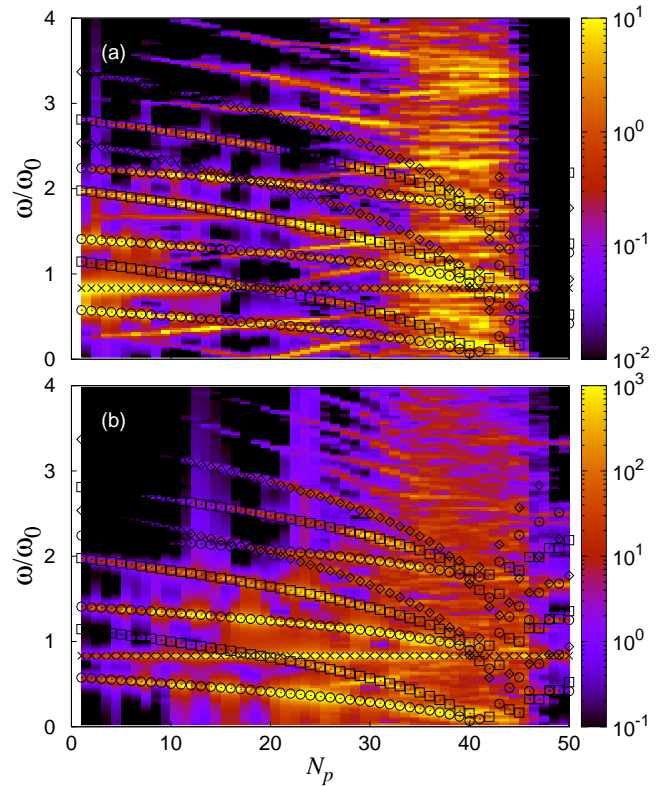}
  \vspace{-0.23cm}
\caption{(Color online) Density plots of (a) $|N(\omega)|^2$ and (b) $|M(\omega)|^2$
as a function of $\omega/\omega_0$ and $N_p$. We also report, as open symbols, 
results for $\omega_1$ through $\omega_9$ (see caption in Fig.~\ref{fig:FT}) ordered 
from bottom to top on the left side of each panel. Note that, following the 
convention in Fig.~\ref{fig:spect}, results involving $E_1$ ($\omega_{1,4,6}$) 
are depicted by circles, $E_2$ ($\omega_{3,5,8}$) by squares, $E_3$ ($\omega_{7,9}$)
by diamonds, and $2\omega'$ ($\omega_{2}$) by crosses.}
 \label{fig:CP}
\end{figure}

In order to illustrate the effect of larger numbers of particles, 
we show in Fig.~\ref{fig:CP} density plots of $|N(\omega)|^2$ 
[Fig.~\ref{fig:CP}(a)] and of $|M(\omega)|^2$ [Fig.~\ref{fig:CP}(b)]
vs $\omega/\omega_0$ and $N_p$. Open symbols depict the values of $\omega_1$
through $\omega_9$, introduced in Fig.~\ref{fig:FT}, as a function of the number 
of particles. Note that for most fillings before the Mott insulator appears 
in the center of the trap, $\omega_{1,4}$, $\omega_2$, and $\omega_5$ are the 
frequencies at which both Fourier transforms have their maximal values. Lattice 
effects are strongest in $|N(\omega)|^2$, where, even for 
the smallest number of particles, frequencies other than $\omega_{1,2,4,5}$ are 
highlighted. Furthermore, one can also see lines of high intensity whose frequencies 
increase with increasing $N_p$. For the cases we could identify (not shown),
they involve combinations of $\omega'$ with $-E_1$ and $-E_2$. None 
of those appear in the analytic solution in the continuum. The results 
in Fig.~\ref{fig:CP}(a) apply to both HCBs and noninteracting fermions to which 
HCBs can be mapped, as their site occupancies are identical.

Overall, the best results for the Fourier analysis are 
obtained for $|M(\omega)|^2$, as shown in Fig.~\ref{fig:CP}(b). 
In that case, most frequencies $\omega_1$ through $\omega_9$ are easily 
identifiable, and lattice effects are the weakest for the lowest fillings, 
where only $\omega_{1,2,4,5}$ are clearly seen in Fig.~\ref{fig:CP}(b) 
[compare with Fig.~\ref{fig:spect}(b)]. When a
Mott insulator is present in the center of the trap (fillings above $N_p=45$), 
one can see that $|N(\omega)|^2$ exhibits almost no response 
(as expected). $|M(\omega)|^2$, on the other hand, exhibits a 
response that is consistent with some of the predictions indicated 
by the open symbols. This supports the view that a Fourier analysis of 
$\ln[m_{k=0}(t)]$, a quantity that behaves very differently for HCBs and fermions 
\cite{rigol_muramatsu_04c,rigol_muramatsu_05c}, is better suited
to study the lowest excitations of the trapped system. 

Since HCBs correspond to the $U/J\rightarrow\infty$ limit of the Bose-Hubbard 
model (where $U$ is the on-site interaction), the Fourier analysis of 
$\ln[m_{k=0}(t)]$ can become a powerful tool to study single-particle 
excitations of Bose-Hubbard-like systems in the presence of a harmonic confinement 
for strong interactions ($U\gg V,J$). As a matter of fact, an exact diagonalization 
analysis of the Bose-Hubbard model in the presence of a harmonic trap presented in 
Ref.~\cite{rey_pupillo_05} showed that, for $U/J>10$, the difference between 
the lowest-energy excitations of soft-core bosons and hard-core bosons scales 
as $J^2/U$, and the latter accurately describes the dynamics of the former. 
As such, we expect that the approach discussed here will be relevant to 
systems with $U/J>10$ and fillings $n\leq 1$ in the center of the trap.

\section{Conclusions}\label{sect:con}

We have shown that the study of the dynamics of site occupancies 
and momentum distribution functions of trapped particles under driven dipole 
oscillations reveals the lowest-energy excitations of the system. The analysis 
of the momentum distribution function was found to provide the best results 
for lattice hard-core bosons, and we expect this to extend to the soft-core case
in the presence of strong interactions.
As opposed to approaches that use lattice modulations, with this approach one does not 
need to probe the system under different driving frequencies.  We have also studied
(not shown) driven systems in which the strength of the confining potential 
is the one that is periodically modulated. In that case, the lowest excitations 
that preserve parity can be determined by studying the Fourier transform of the same
observables as considered here.

\section{Acknowledgments}
This work was supported by the Office of Naval Research (K.H. and M.R.)
and by Department of Energy Grant No.~DE-FG02-05ER46240 (S.H.). 
We thank Aditya Raghavan and David Weiss for useful discussions.


%

\end{document}